\documentclass[twocolumn, showpacs,amsmath,amssymb]{revtex4}
\usepackage{graphicx}
\usepackage{dcolumn}
\usepackage{bm}
\usepackage{amsmath}
\usepackage{amsfonts}
\usepackage{amssymb}
\usepackage{amsthm}
\usepackage{epstopdf}
\usepackage{longtable}
\usepackage{array}
\usepackage{pifont}
\usepackage{latexsym}
\usepackage[latin1]{inputenc}
\usepackage{bm}
\usepackage{color}
\newcolumntype{.}{D{.}{.}{-1}}

\begin{document}

\preprint{APS/123-QED}

\title{Quantum engineering of atomic phase-shifts in optical clocks}

\author{T. Zanon-Willette$^{1,2,3}$\footnote{E-mail address: thomas.zanon@upmc.fr \\}, S. Almonacil$^{4}$, E. de Clercq$^{5}$, A. D. Ludlow$^{6}$, and E. Arimondo$^{7}$}
\affiliation{$^{1}$Sorbonne Universités, UPMC Univ Paris 06, UMR 8112, LERMA, Paris, France}
\affiliation{$^{2}$CNRS, UMR 8112, LERMA, Paris, France}
\affiliation{$^{3}$PSL Research University, Observatoire de Paris, UMR 8112, LERMA, Paris, France}
\affiliation{$^{4}$ Institut d'Optique Graduate School, 2 avenue Augustin Fresnel, 91127 Palaiseau Cedex, France}
\affiliation{$^{5}$LNE-SYRTE, Observatoire de Paris, CNRS, UPMC, 61 avenue de l'Observatoire, 75014 Paris, France}
\affiliation{$^{6}$National Institute of Standards and Technology, 325 Broadway, Boulder, Colorado 80305, USA}
\affiliation{$^{7}$Dipartimento di Fisica "E. Fermi", Universit\`a di Pisa, Lgo. B. Pontecorvo 3, 56122 Pisa, Italy}
\date{\today}

\begin{abstract}
Quantum engineering of time-separated Raman laser pulses in three-level systems is presented to produce an ultra-narrow optical transition in bosonic alkali-earth clocks free from light shifts and with a significantly reduced sensitivity to laser parameter fluctuations. Based on a quantum artificial complex wave-function analytical model, and supported by a full density matrix simulation including a possible residual effect of spontaneous emission from the intermediate state, atomic phase-shifts associated to Ramsey and Hyper-Ramsey two-photon spectroscopy in optical clocks are derived. Various common-mode Raman frequency detunings are found where the frequency shifts from off-resonant states are canceled, while strongly reducing their uncertainties at the $10^{-18}$ level of accuracy.
\end{abstract}

\maketitle

\section{Introduction}

The control and even the cancelation of systematic frequency shifts inherent in atom-light interactions are important tasks for high precision measurement in optical lattice clocks exploiting high quality factors from ultra-narrow transitions~\cite{DereviankoKatori:2011}. For instance, optical clocks avoid a dephasing of the clock states through carefully designed optical traps producing controlled ac Stark shifts of those states~\cite{YeKimbleKatori:2008}. Engineering the phase-shifts which dephase a wave-function is also strongly relevant to a wide range of quantum matter experiments using trapped ions, neutral atoms, and cold molecules \cite{Kajita:2011}. The standard approach for reducing those phase-shifts is the decrease of the probe laser intensity. Continuous progress in the manipulation of the laser-atom/molecule interaction has opened a new direction, quantum state engineering, in the quantum control of atomic/molecular systems \cite{Gibble:2009,Maineult:2012,Hazlett:2013}. For ultracold atoms, quantum engineering leads to the quantum simulation of Hamiltonians describing different physical systems, as in the synthesization of magnetic fields exploiting the coupling between internal and external atomic states~\cite{LinSpielmann:2009,Ketterle:2013,Bloch:2013}. Elsewhere, highly coherent and precisely controlled optical lattice clocks are explored for quantum simulation of many-body spin systems~\cite{MartinReyYe:2013} and optical-clock systems have been proposed to probe the many-body atomic correlation functions~\cite{KnapBlochDemler:2013}.\\
\indent The present work employs the quantum engineering of atom-laser interactions to produce a perfect cancelation of the frequency shifts of a given atomic/molecular level scheme. This cancelation may be applied whenever the atomic wave-function evolution is modified by ac Stark shifts from levels both internal and external to the probed system. By making use of laser pulse sequences shaped in duration, intensity and phase and  modeled by a synthesized Hamiltonian, we easily control these shifts to a challenging $10^{-18}$ level of optical clock accuracy~\cite{LeTargat:2013}. This quantum engineering method allows us to explore various experimental conditions for optical clocks based on alkaline earth atoms, but it can be applied to any other systems dealing with a careful control of the wave-function phase-shift.\\

\indent Our attention is focused on two-photon bosonic optical lattice clocks with a cancelation of the clock shift at the above level of accuracy. One-photon systems such as those using fermionic species in an optical lattice clock have made great progress recently~\cite{LeTargat:2013}. From a metrological perspective the bosonic species of optical lattice clocks have several favorable characteristics over their fermionic counterparts, e.g.~simpler internal structure, distinct collision effects, and suppressed influence of higher-order lattice polarization effects.
However, in order to take advantage of these benefits, a multi-photon interrogation scheme must be able to probe the atomic system without introducing significant Stark effects. The alkaline-earth fermionic species have been also useful for exploring both two and many-body atomic interactions in a well-controlled quantum system, by leveraging the clock transition as a high-precision measurement tool~\cite{MartinReyYe:2013}. The same can be true of the bosonic species. However, to make such a measurement would require high precision frequency measurements also in the bosonic systems by simultaneously activating the forbidden transition and canceling the systematic frequency shifts~\cite{Ludlow:2014}.\\
\indent For two-photon optical clocks, the ac Stark shifts~\cite{Liao:1975,CohenTannoudji:1977} are induced by a large number of off-resonant driven transitions, making their suppression very difficult to realize in a simple manner. Cancelation of the frequency shifts in those optical clocks using pulsed Electromagnetically Induced Transparency and Raman (EIT-Raman) spectroscopic interrogation was explored in~\cite{ZanonWillette:2006,Yoon:2007}. Alternatively, frequency-shifts of one-photon clock transitions were compensated and the uncertainty of the frequency measurement strongly reduced using so-called Hyper-Ramsey spectroscopy with two Ramsey pulses of different areas, frequencies and phases~\cite{Yudin:2010}. Refs.~\cite{Taichenachev:2009,Tabatchikovaa:2013} pointed out that a laser frequency step applied during that pulse sequence cancels the residual light-shift and an additional echo pulse compensates both the dephasing and the uncontrolled variations of the pulse area, as in a recent  $^{171}$Yb$^+$ ion clock experiment ~\cite{Huntemann:2012}. Within our quantum engineering approach we introduce here generalized Hyper-Raman-Ramsey (HRR) techniques for two-photon optical clocks and derive precise conditions required to operate at the $10^{-18}$ level of accuracy and stability. To reach this level of performance, the following criteria are sought: i) the clock phase-shift from ac Stark effects is eliminated, ii) the shift cancelation is stable against fluctuations in the laser parameters; iii) the Ramsey fringe contrast is maximized; iv) that contrast is obtained at the unperturbed clock frequency. We satisfy simultaneously all these targets with the generalized three-level HRR techniques.

\indent We show that all the light shift contributions to the clock transition, from internal and external states, can be canceled by operating the excitation lasers at magic detuning, more precisely a "magic" common mode detuning for the Raman configuration. The light shift compensation is based on control of the Bloch vector phase in the equatorial plane before the last pulse in the HRR sequence. A properly oriented vector combined with a well chosen pulse area produces full occupation of the final state, and this condition is ideally realized within the standard Ramsey sequence~\cite{Ramsey:1950}. However, in presence of light shift and relaxation processes, the Bloch vector is not properly aligned and the final pulse cannot produce the required state. Those imperfections may be compensated by a final pulse with a tuned pulse area, while the incorrect phase of the wave-function is controlled by the echo pulse, and eventually by a laser phase reversal.\\

\indent Our theoretical approach is based on a non hermitian evolution of the atomic wave-function whose phase is modified by the ac Stark shifts from levels internal and external to the probed system. This approach enables the time-dependent solution of the atomic phase-shift to be computed in analytical form, making possible a detailed exploration of laser parameters which cancel the frequency shifts.
\begin{figure}[!!b]
\centering%
\resizebox{8.cm}{!}{\includegraphics[angle=-90]{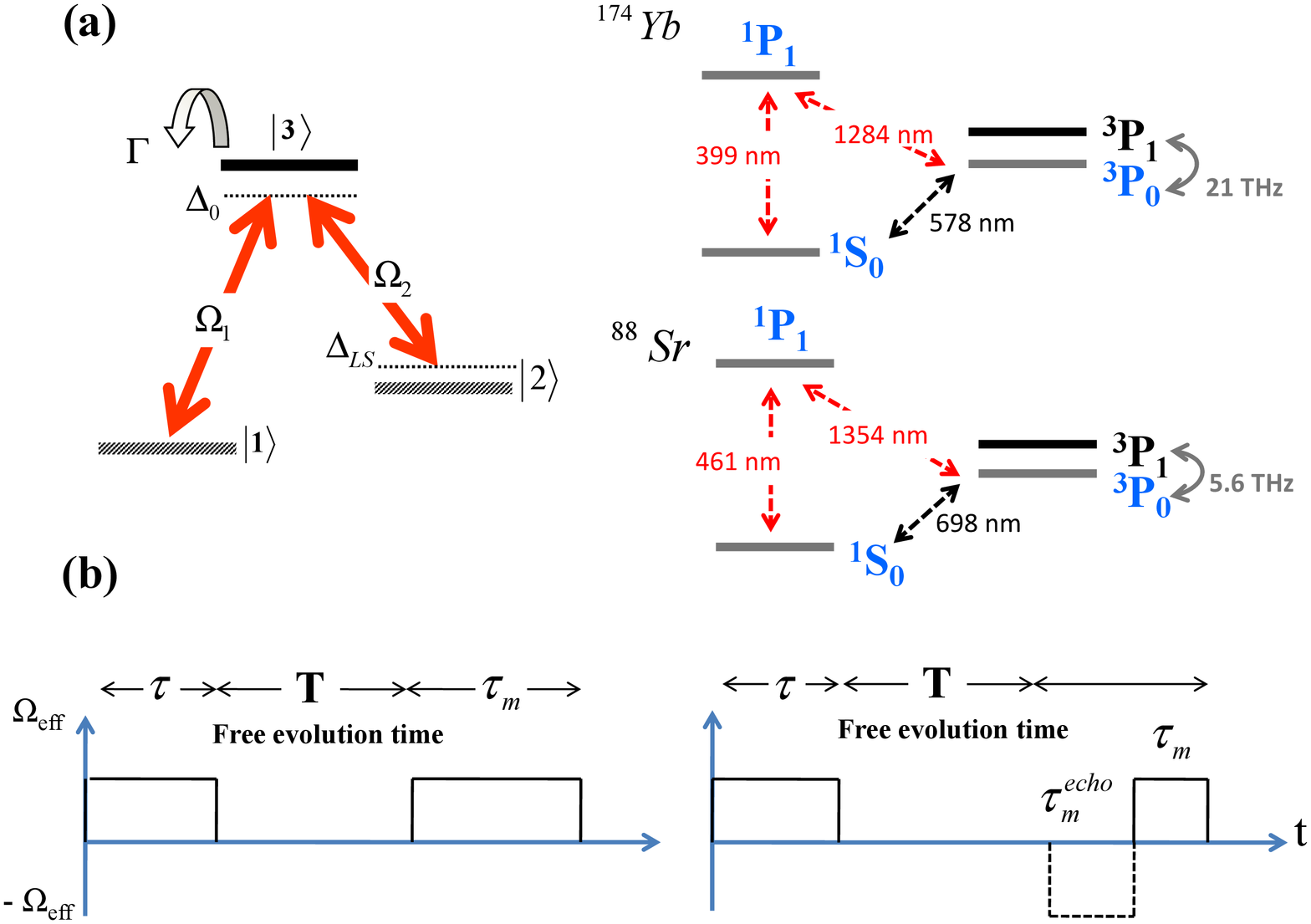}}
\caption{(color online) (a) Three-level atomic system configuration for spectroscopy of a forbidden atomic transition in $^{174}$Yb and $^{88}$Sr optical lattice clocks. Atomic parameters are described in text. (b) EIT-Raman laser pulses in a Ramsey-type (on,off,on) sequence with interrogation times $\tau$, $\tau_{m}^{echo}$ (if inserting an echo pulse) and $\tau_m$, respectively, and free evolution at the clock frequency for a time $T$.}
\label{lambda-scheme}
\end{figure}
The three-level system and atomic parameters for an homogeneous medium are presented in Sec.~II.
We develop in Sec.~III a wave-function model including radiative correction which is compared to the density matrix results. We derive the key informations on the phase-shift of the wave-function particle. In Sect.~IV, we synthesize a general analytical phase-shift expression between clock states leading to a frequency-shift of the central fringe. In Sec.~V, we develop a numerical analysis of the external light-shift contribution derived from the dynamic polarizability calculations. Finally Sec.~VI analyzes the resulting Hyper-Raman Ramsey fringes produced with highly-detuned two-photon pulses separated in time and the cancelation of the external light shift at particular magic detunings from the intermediate state.

\section{Atomic parameters for a three-level system}

\indent We examine three-level optical bosonic clocks for $^{88}$Sr and $^{174}$Yb with the level structure of Fig.~\ref{lambda-scheme}(a). The doubly forbidden optical clock transition is driven by a two-photon transition between atomic states $|1\rangle\equiv |^1S_0\rangle$ and $|2\rangle\equiv |^3P_0\rangle$ via the $|3\rangle\equiv|^1P_1\rangle$ off-resonantly excited state. The $|3\rangle\to |2\rangle$ transition is driven by laser induced magnetic coupling~\cite{SantraYe:2005}. Our quantum engineering approach could also be applied to the magnetically induced spectroscopy scheme where a static magnetic-field induces the $|3\rangle\to|2\rangle$ transition as in~\cite{Taichenachev:2006,Barber:2006,Baillard:2007,Akatsuka:2010}. The present work explores different laser pulsed excitation schemes, as shown generally in Fig.~\ref{lambda-scheme}(b), with a free-evolution time $T$ and (eventually) an echo pulse duration $\tau_{m}^{echo}$ with a laser phase reversal taking place between the initial and final interrogation times, $\tau$ and $\tau_m$ respectively. The clock transition is probed via detection of the $^{1}S_{0}$ or $^{3}P_{0}$ populations as in ref~\cite{ZanonWillette:2006}.\\

\indent Within our model, the Rabi frequencies $\Omega_{i} (i=1,2)$ are defined by electric and magnetic dipole couplings. The laser detunings are introduced as $\Delta_{1}=\Delta_{0}+\eta_{1}$ and $\Delta_{2}=\Delta_{0}-\delta+\eta_{2}$, where $\Delta_0$ is the common mode frequency detuning from the excited state, $\delta$ is the Raman clock frequency detuning. $\eta_i$, the light-shift correction induced by external levels for the $(i=1,2)$ corresponding transition,  produces a  correction $\Delta_{ext}=\eta_{2}-\eta_{1}$ to the field-free clock transition. That transition also experiences an internal shift leading to the $\Delta_{LS}$ total shift listed in the first line of the Tab.~\ref{parameters}. The effective complex Rabi frequency $\Omega_{eff}$ of Tab.~\ref{parameters} determines the two-photon  coupling between $|1\rangle$ and $|2\rangle$. The spontaneous emission rate from $|3\rangle$ is $\Gamma$, optical relaxations are $\gamma_{1}=\gamma_{2}=\Gamma/2$. Spontaneous decays to external levels may be easily included into the model.\\
\indent In order to perform an accurate analysis of the population and coherence temporal dynamics, we adopt two points of views in a bare basis: a density matrix formalism and a complex wave function approach. The numerical solution of the density matrix equations of ref.~\cite{Zanon-Willette:2011} allows us to examine with high accuracy the lineshapes and shifts of the clock transition as in \cite{Taichenachev:2009,Yudin:2010,ZanonWillette:2006,Huntemann:2012}.\\

\section{Wave-function analysis including radiative correction}

\begin{table}[t!!]
\centering
\begin{tabular}{|c|c|c|}
\hline
\hline
& Two-photon \\
\hline
$\Delta_{LS}$ &  $\Delta_{ext}+(\Omega_{2}^{2}-\Omega_{1}^{2})\frac{\Delta_{0}}{\Delta_{0}^{2}+\Gamma^{2}/4}$  \\
\hline
$\Delta_{eff}$ & $\Delta_{LS}-i(\Omega_{2}^{2}-\Omega_{1}^{2})\frac{\Gamma/2}{\Delta_{0}^{2}+\Gamma^{2}/4}$   \\
\hline
$\Omega_{eff}$ & $2\Omega_{1}\Omega_{2}\frac{\Delta_{0}-i\Gamma/2}{\Delta_{0}^{2}+\Gamma^{2}/4}$   \\
\hline
\hline
\end{tabular}
\caption{Two-photon wave-function parameters including radiative correction.}
\label{parameters}
\end{table}
This section reports the formalism presented in \cite{ZanonWillette:2006,Zanon-Willette:2005,Yoon:2007} determining the phase accumulated by the
atomic wave-function including all the ac Stark shifts. It is based on an effective non hermitian two-level Hamiltonian \cite{Carmichael:1993} describing the $|\Psi(t)\rangle$ superposition of the $|1\rangle,|2\rangle$ clock states
\begin{equation}
|\Psi(t)\rangle= c_\alpha(t)|1\rangle+c_\beta(t)|2\rangle.
\end{equation}
This effective wave-function model includes complex energies for open quantum systems.
The model, although unadapted to conserve atomic population, is still valid as long as pulses are applied with short interaction times avoiding any cw stationary regime. We therefore expect spontaneous emission to have only a perturbative effect on the dynamics and choose to work in terms of the state vector $|\Psi(t)\rangle$ and its Schr\"odinger equation. The effects of the decoherence emission may be included by using a $\Delta_{eff}$ complex Raman detuning with a term $-i\Gamma/2$ associated to spontaneous emission from the intermediate state inside the $\Delta_0$ common-mode detuning. This replacement is computationally simpler than solving a full master equation. The norm of the state vector calculated using the previous assumption is not constant, since after introducing a complex detuning, the Hamiltonian is no longer Hermitian. By making this replacement, we are still taking a conservative approach in the sense that this reduced-norm state vector is always less than or at least
equal to the true calculation using the master equation.\\
\indent The $|\Psi(t)\rangle$ evolution is driven by the Hamiltonian $H$
\begin{equation}
\begin{array}{l}
\frac{H}{\hbar}=\left(\begin{array}{cccccc} &0&\Omega_{eff}\\
&\Omega_{eff}&\delta_{eff}
\end{array}
\right)
\end{array}
\end{equation}
where $\delta_{eff}=-\delta+\Delta_{eff}$ and the effective complex Rabi frequency $\Omega_{eff}$ of Tab.~\ref{parameters} determines the two-photon coupling between $|1\rangle$ and $|2\rangle$ \cite{comments}.
The $\Delta_{eff}$ detuning includes contribution from the three-level itself and the $\Delta_{ext}$ external light-shift contribution from others levels on the clock detuning.\\
\indent Using the solution of the Schrödinger's equation, we write for the $c_{1,2}(\theta)$ transition amplitudes
\begin{equation}
\begin{split}
\left(%
\begin{array}{c}
c_{\alpha}(\theta)
\\
c_{\beta}(\theta)
\\
\end{array}%
\right)
=\chi(\theta)\cdot
\left(%
\begin{array}{cc}
M_{+}(\theta)
&M_{\dagger}(\theta)
\\
M_{\dagger}(\theta)
& M_{-}(\theta) \\
\end{array}%
\right) \cdot
\left(%
\begin{array}{c}
c_{\alpha}(0) \\
c_{\beta}(0) \\
\end{array}%
\right)
\end{split}
\end{equation}
including a phase factor of the form
\begin{equation}
\begin{split}
\chi(\theta)=\exp\left[-i\left(\frac{\Delta_{0}-i\Gamma/2}{\Delta_{0}^2+\frac{\Gamma^{2}}{4}}(\Omega_2^2+\Omega_1^2)-\delta+\Delta_{ext}\right)\frac{t}{2}\right]
\end{split}
\label{Ramsey-phase-factor}
\end{equation}
where the wave-function evolution driven by the pulse area $\theta$ is determined by the following complex 2x2 interaction matrix $M$ \cite{comments}:
\begin{equation}
\begin{split}
M(\theta)&=\left(
\begin{array}{cc}
M_{+}(\theta) &M_{\dagger}(\theta) \\
M_{\dagger}(\theta) & M_{-}(\theta) \\
\end{array}%
\right)\\
&=\left(%
\begin{array}{cc}
\cos(\theta)+i\frac{\delta_{eff}}{\omega}\sin(\theta)&-i\frac{2\Omega_{eff}}{\omega}\sin(\theta) \\
-i\frac{2\Omega_{eff}}{\omega}\sin(\theta)&\cos(\theta)-i\frac{\delta_{eff}}{\omega}\sin(\theta) \\
\end{array}
\right).
\end{split}
\label{matrix}
\end{equation}
with $\omega^{2}=\delta_{eff}^{2}+\Omega_{eff}^{2}$.
The pulsed excitation is written as a product of different matrices $M(\theta_{i,j,k})$ and a free evolution without laser light during a Ramsey time T described as a composite pulsed sequence of three different pulse areas $\theta_{i}-T-\theta_{j}-\theta_{k}$. Pulse areas are defined by $\theta_{i,j,k}=\omega\tau_{i,j,k}/2$ with $\tau_{i,j,k}\equiv\tau,\tau_{m}^{echo},\tau_{m}$.
The final Hyper-Ramsey expression is a product of matrices which depends on initials
conditions $c_{\alpha}(0)$ and $c_{\beta}(0)$ as \cite{Zanon-Willette:2005}:
\begin{widetext}
\begin{equation}
\begin{split}
&\left(%
\begin{array}{c}
c_{\alpha}(\theta_{i},T,\theta_{j},\theta_{k}) \\
c_{\beta}(\theta_{i},T,\theta_{j},\theta_{k}) \\
\end{array}%
\right)=\chi(\theta_{i},\theta_{j},\theta_{k})\cdot\left(%
\begin{array}{cc}
\begin{split}& M_{+}(\theta_{i})M_{+}(\theta_{j},\theta_{k})\\+&M_{\dagger}(\theta_{j},\theta_{k})M_{\dagger}(\theta_{i})exp\left[i\delta T\right]\end{split}& \begin{split}&M_{\dagger}(\theta_{i})M_{+}(\theta_{j},\theta_{k})\\+&M_{\dagger}(\theta_{j},\theta_{k})M_{-}(\theta_{i})exp\left[i\delta T\right]\end{split}
\\\\
\begin{split}& M_{\dagger}(\theta_{k},\theta_{j})M_{+}(\theta_{i})\\+&M_{\dagger}(\theta_{i})M_{-}(\theta_{j},\theta_{k})exp\left[i\delta T\right]\end{split}& \begin{split}&M_{\dagger}(\theta_{k},\theta_{j})M_{\dagger}(\theta_{i})\\+&M_{-}(\theta_{j},\theta_{k})M_{-}(\theta_{i})exp\left[i\delta T\right]\end{split} \\
\end{array}%
\right)\cdot\left(%
\begin{array}{c}
c_{\alpha}(0) \\
c_{\beta}(0) \\
\end{array}%
\right)
\end{split}
\end{equation}
\end{widetext}
where $\chi(\theta_{i},\theta_{j},\theta_{k})=\chi(\theta_{i})\chi(\theta_{j})\chi(\theta_{k})$ and
\begin{equation}
\begin{split}
M_{+}(\theta_{j},\theta_{k})&=M_{+}(\theta_{j})M_{+}(\theta_{k})+M_{\dagger}(\theta_{k})M_{\dagger}(\theta_{j})\\
M_{-}(\theta_{j},\theta_{k})&=M_{\dagger}(\theta_{j})M_{\dagger}(\theta_{k})+M_{-}(\theta_{k})M_{-}(\theta_{j})\\
M_{\dagger}(\theta_{j},\theta_{k})&= M_{\dagger}(\theta_{j})M_{+}(\theta_{k})+M_{\dagger}(\theta_{k})M_{-}(\theta_{j})\\
M_{\dagger}(\theta_{k},\theta_{j})&= M_{\dagger}(\theta_{k})M_{+}(\theta_{j})+M_{\dagger}(\theta_{j})M_{-}(\theta_{k})
\end{split}
\end{equation}

\section{Synthesized phase-shift for light shift control}

\indent A key point for clock precision is the quantum control of the $\Delta_{LS}$ shift, with the frequency shift of the central Ramsey clock fringe approximatively given by $\delta\nu\sim\Delta_{LS}\times\tau/T$ for $\tau\approx\tau_m$~\cite{Ramsey:1950}. We show that all the light shift contributions to the clock transition, from internal and external states, can be canceled by operating the excitation lasers at a magic common mode detuning $\Delta_0^{m}$. The optical clock applications of a synthesized shift require simultaneous cancelation and stability of the phase shift at the magic detuning, as expressed by the conditions
\begin{subequations}
\begin{align}
\left[\delta \nu\right]_{\Delta_0^m}&=0\\
\left[\frac{\partial\delta \nu}{\partial \Delta_0}\right]_{\Delta_0^m}&=0\\
\left[\frac{\partial^{2}\delta \nu}{\partial\Delta_0^{2}}\right]_{\Delta_0^m}&=0.
\end{align}
\label{conditions}
\end{subequations}
In analogy with quantum simulations using ultracold atoms~\cite{Bloch:2012}, our synthesized frequency shift may be described through an effective Hamiltonian determining the final atomic state at the end of the full pulse sequence.
The final expression is an Hyper-Ramsey complex amplitude including initial atomic/molecular state preparation.
We are then able to explicit transition probabilities with the following general form:
\begin{equation}
\begin{split}
P_{\alpha\beta}&=c_{\alpha}(\theta_{i},T,\theta_{j},\theta_{k})c_{\beta}^{*}(\theta_{i},T,\theta_{j},\theta_{k})\\
&=|\alpha_{\alpha\beta}|^{2}\left|1+\beta_{\alpha\beta}e^{i(\delta T+\Phi_{\alpha\beta})}\right|^{2}
\label{probabilities}
\end{split}
\end{equation}
using the $\alpha,\beta=1,2$ for state label. The complex term $\Phi_{\alpha\beta}$ represents the atomic phase-shift accumulated by the wave-function during the laser interrogation sequence. The wave-function expression for atomic population $P_{22}$ of our effective two-level system is then:
\begin{equation}
\small{
\begin{split}
\alpha_{22}=&\left[M_{+}(\theta_{i})c_{1}(0)+M_{\dagger}(\theta_{i})c_{2}(0)\right]\cdot M_{\dagger}(\theta_{k},\theta_{j})\\
\beta_{22}e^{i\Phi_{22}}=&\left[\frac{M_{\dagger}(\theta_{i})c_{1}(0)+M_{-}(\theta_{i})c_{2}(0)}{M_{+}(\theta_{i})c_{1}(0)+M_{\dagger}(\theta_{i})c_{2}(0)}\right]\cdot\frac{M_{-}(\theta_{j},\theta_{k})}{M_{\dagger}(\theta_{k},\theta_{j})}.
\end{split}}
\label{population_22}
\end{equation}
\indent The wave-function formalism adopted here with complex state energies leads to atomic phase-shifts extracted from population transition probabilities.
The population transition probability $P_{22}$ is used to evaluate both lineshape, population transfer and frequency-shift affecting the clock transition and is compared to a numerical density matrix calculation describing dynamics of a closed three-level system.
For the case of an highly EIT/Raman detuned regime from the intermediate state where spontaneous emission can be neglected, including a $\pi$ phase reversion of one laser field,
we provide an analytical form of the associated phase-shift. Starting from an initial condition $c_{1}(0)=1$ and $c_{2}(0)=0$, the phase-shift $\Phi(^{3}P_{0})\equiv\Phi_{22}$ is given by:
\begin{equation}
\begin{split}
\Phi(^{3}P_{0})&=Arg\left[\frac{M_{\dagger}(\theta_{i})}{M_{+}(\theta_{i})}\cdot\frac{M_{-}(\theta_{j},\theta_{k})}
{M_{\dagger}(\theta_{k},\theta_{j})}\right].
\label{Population_22-shift_1}
\end{split}
\end{equation}
\indent Within the limit of $\theta_{j}\mapsto0$, we recover the phase-shift expression established in ref.~\cite{ZanonWillette:2006} for an EIT/Raman excitation and given by:
\begin{equation}
\begin{split}
\Phi(^{3}P_{0})=Arg\left[\frac{M_{\dagger}(\theta_{i})}{M_{+}(\theta_{i})}\cdot\frac{M_{-}(\theta_{k})}{M_{\dagger}(\theta_{k})}\right].
\label{Population_22-shift}
\end{split}
\end{equation}
The phase-shift $\Phi(^{3}P_{0})$ determines the optical lattice clock shift $\delta \nu(\Delta_{0})$ measured on the central fringe in the two-photon HRR spectroscopy.
When $T>>\tau,\tau_{m}^{echo},\tau_{m}$, that shift is given by
\begin{equation}
\delta\nu(\Delta_{0})\approx-\frac{\Phi(^{3}P_{0})}{2\pi T}.
\label{frequency-shift}
\end{equation}
\indent In the $\Delta_0\gg\Gamma$ regime, the role of spontaneous emission is strongly reduced and the population transfer to the excited state is negligible.  In this case, the clock phase-shift depends only on pulse area. The full expression of the $\Phi(^{3}P_{0})$ phase-shift corresponding to a composite sequence of three different laser
pulse areas $\theta_{i},\theta_{j},\theta_{k}$ including a $\pi$ phase reversion during the second pulse is:
\begin{widetext}
\begin{equation}
\Phi(^{3}P_{0})=-\arctan\left[\frac{\frac{\Delta_{eff}}{\omega}\left(\frac{\tan\theta_j
+\tan\theta_k}{1+\frac{\Omega_{eff}^2-\Delta_{eff}^2}{\omega^2}\tan\theta_j\tan\theta_k}
+\frac{\tan\theta_i-2\frac{\tan\theta_j\tan\theta_k}{\tan\theta_j-\tan\theta_k}}{1+2\frac{\Delta_{eff}^{2}}{\omega^{2}}
\frac{\tan\theta_i\tan\theta_j\tan\theta_k}{\tan\theta_j-\tan\theta_k}}\right)}{1-\left(\frac{\Delta_{eff}}{\omega}\right)^{2}\frac{\tan\theta_j
+\tan\theta_k}{1+\frac{\Omega_{eff}^2-\Delta_{eff}^2}{\omega^2}\tan\theta_j\tan\theta_k}\cdot\frac{\tan\theta_i-2\frac{\tan\theta_j\tan\theta_k}{\tan\theta_j-\tan\theta_k}}{1+2\frac{\Delta_{eff}^{2}}{\omega^{2}}
\frac{\tan\theta_i\tan\theta_j\tan\theta_k}{\tan\theta_j-\tan\theta_k}}}\right]
\label{HER-phase-shift}
\end{equation}
\end{widetext}
When the second pulse area is $\theta_{j}\mapsto0$ and $\theta_{i}\neq\theta_{k}$, we recover the full expression of the Hyper-Ramsey phase-shift expression derived perturbatively in \cite{Yudin:2010} for a two-level system:
\begin{subequations}
\small{
\begin{align}
\Phi(^{3}P_{0})&=-\arctan\left[\frac{\frac{\Delta_{eff}}{\omega}\left(\tan\theta_i+\tan\theta_k\right)}{1-\left(\frac{\Delta_{eff}}{\omega}\right)^2\tan\theta_i\tan\theta_k}\right]\\
&=-\arctan\left[\frac{\Delta_{eff}}{\omega}\tan\theta_i\right]-\arctan\left[\frac{\Delta_{eff}}{\omega}\tan\theta_k\right]
\end{align}}
\label{Hyper-Raman-frequency-shift}
\end{subequations}
Based on ref~\cite{Abramowitz:1968}, equations Eq.~\ref{Hyper-Raman-frequency-shift}(a) and Eq.~\ref{Hyper-Raman-frequency-shift}(b) are fully equivalent.
Note that Eq.~(\ref{HER-phase-shift}) also exhibits this remarkable mathematical form.
For the well-known Ramsey configuration \cite{Ramsey:1950} where $\theta_{i}=\theta_{k}=\theta$, that shift is
\begin{subequations}
\small{
\begin{align}
\Phi(^{3}P_{0})&=-\arctan\left[\frac{2\frac{\Delta_{eff}}{\omega }\tan\theta}{1-\left(\frac{\Delta_{eff}}{\omega}\right)^2\tan\theta^2}\right]\\
&=-2\arctan\left[\frac{\Delta_{eff}}{\omega}\tan\theta\right]
\end{align}}
\label{Ramsey-clock-state-shift}
\end{subequations}
\indent From a geometrical point of view, this Ramsey phase-shift is exactly two times the Eulerian angle accumulated by a Bloch vector projection of rotating components in the complex plane using a two dimensional Cauley-Klein representation of the spin 1/2 rotational group~\cite{Jaynes:1955,Schoemaker:1978}.\\

\section{Light shift numerical analysis}

\indent The light shift $\Delta_{LS}$ of the clock states containing contributions from the three-level system itself and  the $\Delta_{ext}$ contribution is written as
\begin{equation}
\begin{split}
\Delta_{LS}&=\left(\Delta^{res}_2-\Delta^{res}_1\right)+\Delta_{ext}\\
&=\left(\Delta^{res}_2-\Delta^{res}_1\right)+ \left(\eta_2-\eta_1\right),
\end{split}
\label{externalshift}
\end{equation}
where $\eta_i$ defines the external contribution from the $i$ level.
The near-resonant light shift may be written as
\begin{equation}
\begin{split}
\Delta^{res}_i&=\frac{\Delta_{0}}{\Delta_{0}^{2}+\Gamma^{2}/4}\Omega_{i}^{2}\\
&=\frac{\Delta_{0}}{\Delta_{0}^{2}+\Gamma^{2}/4}\left(\frac{|\langle i|{\bf d}|3\rangle|}{\hbar}\right)^2\left(\frac{E_{L,i}}{2}\right)^2.
\end{split}
\label{resonantshift}
\end{equation}
with $(i=1,2)$. Here we have introduced the $E_{L,i}$ electric field amplitude for the $i$ laser field, and  the $\langle i|{\bf d}|3\rangle$ ($\langle i|{\bf m}|3\rangle$) electric (magnetic) dipole operator between the $|i\rangle$ state and the $|3\rangle=|n^1P_0\rangle$ intermediate state with $n=(5,6)$ for Sr and Yb, respectively.\\
\begin{table*}
\centering
\begin{tabular}{|c|c|c|c|c|c|c|c|}
\hline
Atom &  $|1\rangle$ or $|2\rangle$ & $|3\rangle$ &$\lambda$ [nm] &   $\omega/2\pi$   &  $\langle 1||{\bf d}||3\rangle$ or $\langle 2||{\bf m}||3\rangle$ [a.u.]  &$\tilde{\alpha_i}$  [a.u.] &$\eta_i$ [mHz/[mW/cm$^2$)] \\
\hline
Sr  & $|1\rangle =|5^1S_0\rangle$ &$|5^1P_0\rangle$& 461 & 650.5 & $\langle1||{\bf d}||3\rangle=$5.248(2)~\cite{YasudaKatori:2006}&48.5&-2.27\\
\hline
Sr  & $|1\rangle =|5^1S_0\rangle$ &$|5^1P_0\rangle$& 1354 & 221.2 & $\langle1||{\bf d}||3\rangle=$5.248(2)~\cite{YasudaKatori:2006}&213.6&-10.01\\
\hline
Sr & $|2\rangle =|5^3P_0\rangle$  &$|5^1P_0\rangle$& 461 & 650.6 & $\langle2||{\bf m}||3\rangle$=0.022$\mu_B$~\cite{SantraYe:2005}&-1220.9&57.22\\
\hline
Sr & $|2\rangle =|5^3P_0\rangle$  &$|5^1P_0\rangle$& 1354 & 221.2 & $\langle2||{\bf m}||3\rangle$=0.022$\mu_B$~\cite{SantraYe:2005}&61.8&-2.90\\
\hline
Yb & $|1\rangle =|6^1S_0\rangle$ &$|6^1P_0\rangle$&  399& 751.5 & $\langle1||{\bf d}||3\rangle$=4.148(2)~\cite{Takasu:2004}&33.0&-1.55\\
\hline
Yb & $|1\rangle =|6^1S_0\rangle$ &$|6^1P_0\rangle$&  1284& 233.2 & $\langle1||{\bf d}||3\rangle$=4.148(2)~\cite{Takasu:2004}&152.9&-7.14\\
\hline
Yb & $|2\rangle =|6^3P_0\rangle$ & $|6^1P_0\rangle$& 399 & 751.5 & $\langle2||{\bf m}||3\rangle$=0.16~\cite{Beloy:2014}&-129.1&6.05\\
\hline
Yb & $|2\rangle =|6^3P_0\rangle$ & $|6^1P_0\rangle$& 1284 & 233.2& $\langle2||{\bf m}||3\rangle$=0.16~\cite{Beloy:2014}&-944.4&44.26\\
\hline
\end{tabular}
\caption{Sr and Yb atomic levels for the two-photon clock  investigated in the present work. The coupling to the $|3\rangle=|$nsnp $^1P_1\rangle$ state is either electric dipole or magnetic dipole transition. For the $i=(1,2)$ level the $\alpha_i$ external light shift contribution is in atomic units  in the column before the last one, with $\eta_i$ in Hz/(mW/cm$^2$) in the last one.}
\label{polarizabilities-parameters}
\end{table*}
\indent The evaluation of the external contribution requires the dipole moment matrix elements and the energy position for all the excited states. It should be however noticed that the  above
$\Delta_{LS}$ quantity is also calculated for determining the magic wavelength in the design of optical traps producing controlled Sr or Yb frequency shifts. We use the definition of refs.~\cite{DzubaDerevianko:2010,DereviankoKatori:2011} for the $\alpha_i(\omega)$ dynamic polarizabilities of the $i=1,2$ states associated to a laser  field at angular frequency $\omega$
\begin{equation}
\small{
\alpha_i(\omega)=\sum_n\frac{|\langle i||{\bf d}||n\rangle|^2}{3}\left[\frac{1}{E_n-E_i-\hbar \omega}+\frac{1}{E_n-E_i+\hbar \omega}\right]},
\label{alphas}
\end{equation}
with ${\bf d}$ again the electric dipole operator (or the ${\bf m}$ magnetic dipole moment for the electric dipole forbidden transitions) and the summation over a complete set of atomic states. We neglect in this analysis higher-order dynamic polarizabilities. Using the above polarizabilities, the light shift of the $|1\rangle-|2\rangle$ transition is
\begin{equation}
\small{
\Delta_{LS}(\omega)=\alpha_1(\omega)\left(\frac{E_{L,1}}{2}\right)^2-\alpha_2(\omega)\left(\frac{E_{L,2}}{2}\right)^2}
\label{polarizability}
\end{equation}
For each $(i=1,2)$ state the sum of Eq.~\eqref{alphas} includes only one near-resonant light shift, while all the remaining ones represent the external light shift. Within that sum and following Eqs.~\eqref{externalshift} and \eqref{resonantshift}, we separate the resonant contribution from the remaining ones denoted by $\tilde{\alpha}_i$
\begin{equation}
\small{
\alpha_i(\omega)=\tilde{\alpha}_i(\omega)+\frac{\Delta_{0}}{\Delta_{0}^{2}+\Gamma^{2}/4}\left(\frac{|\langle i|{\bf d}|3\rangle|}{\hbar}\right)^2},
\label{alphasplitted}
\end{equation}
where
\begin{equation}
\small{
\begin{split}
\tilde{\alpha}_i(\omega)=&\sum_{n \ne 3}\frac{|\langle i||{\bf d}||n\rangle|^2}{3}\left[\frac{1}{E_n-E_i-\hbar \omega}+\frac{1}{E_n-E_i+\hbar \omega}\right]\\
&+\frac{|\langle i||{\bf d}||3\rangle|^2}{3}\frac{1}{E_3-E_i+\hbar \omega}.
\end{split}}
\label{alphatilde}
\end{equation}
\indent The combination of Eqs.~\eqref{externalshift}, \eqref{polarizability} and  \eqref{alphatilde} allow us to derive the $\Delta^{ext}_i$ light shift contribution. We have calculated from the analysis reported in ref.~\cite{Safronova:2012,Safronova:2013} the $\tilde{\alpha}_i$ external levels dynamic polarizabilities at the laser frequencies required for the EIT-Raman and Hyper-Ramsey schemes investigated in the present work, as reported in Tab.~\ref{polarizabilities-parameters}. The $\eta_i$ light shift contributions of the external levels in the last column of the Table are obtained from the $\tilde{\alpha_i}$ using the conversion factor of ref.~\cite{DereviankoKatori:2011}. The $\eta_i$ light shifts are computed in mHz for a laser intensity of 1 mW/cm$^2$.

\section{Application to bosonic optical lattice clocks}

\indent We have applied the density matrix and complex wave-function approaches to the light shift control for several HRR clock interrogation parameters, the $\Delta_{ext}$ external shift contributions and the magic detunings are reported in Tab.~\ref{Magic-fields-table}.
\begin{table}[!!b]
\centering%
\begin{tabular}{cccccc}
\hline
\hline
& &  \textbf{$^{88}$Sr} & & \\
 $I_{1}$ (mW/cm$^{2}$)& 0.066 & 0.016 & 4.2 & 160 & 640 \\
$I_{2}$ (W/cm$^{2}$)& 9.4 &  2.35 &  37.4 & 57.2 & 229  \\
 $\Delta_{ext}/2\pi$ (Hz)& 66 & 17 & 267  &  417 & 1667 \\
 $(\tau,\tau_{m})$ (s)& ($\frac{3}{16}$,$\frac{9}{16}$) & ($\frac{3}{4}$,$\frac{9}{4}$) & ($\frac{3}{16}$,$\frac{9}{16}$) &  ($\frac{3}{5}$,$\frac{9}{5}$) & ($\frac{3}{20}$,$\frac{9}{20}$)  \\
$\Delta_{0}^{m}/2\pi$ (GHz) & 11.2 &  11.2 & 179  & 4370 & 4370 \\
\hline
\hline
& &   \textbf{$^{174}$Yb} &  &  \\
 $I_{1}$ (mW/cm$^{2}$)  & 0.78 & 0.2 & 12.5  &  122 & 487 \\
$I_{2}$ (W/cm$^{2}$) & 1.3 & 0.32 & 5.2  & 8.1 & 32.3 \\
 $\Delta_{ext}/2\pi$ (Hz) & 66  & 17 & 267  & 417 & 1667 \\
 $(\tau,\tau_{m})$ (s) & ($\frac{3}{32}$,$\frac{9}{32}$) & ($\frac{3}{8}$,$\frac{9}{8}$) &  ($\frac{3}{64}$,$\frac{9}{64}$) & ($\frac{3}{40}$,$\frac{9}{40}$) & ($\frac{3}{160}$,$\frac{9}{160}$) \\
$\Delta_{0}^{m}/2\pi$ (GHz) & 82 & 82 & 326.7 & 2037 & 2037 \\
\hline
\hline
\end{tabular}
\caption{Raman interaction parameters including laser intensities $I_{1},I_{2}$ and pulse durations $(\tau,\tau_{m})$, at fixed $T=5$~s, for $^{88}$Sr and $^{174}$Yb two-photon interrogation leading to the listed $\Delta_{0}^m$ magic detuning.}
\label{Magic-fields-table}
\end{table}
\begin{figure}[!!t]
\centering
\resizebox{8.5cm}{!}{\includegraphics[angle=0]{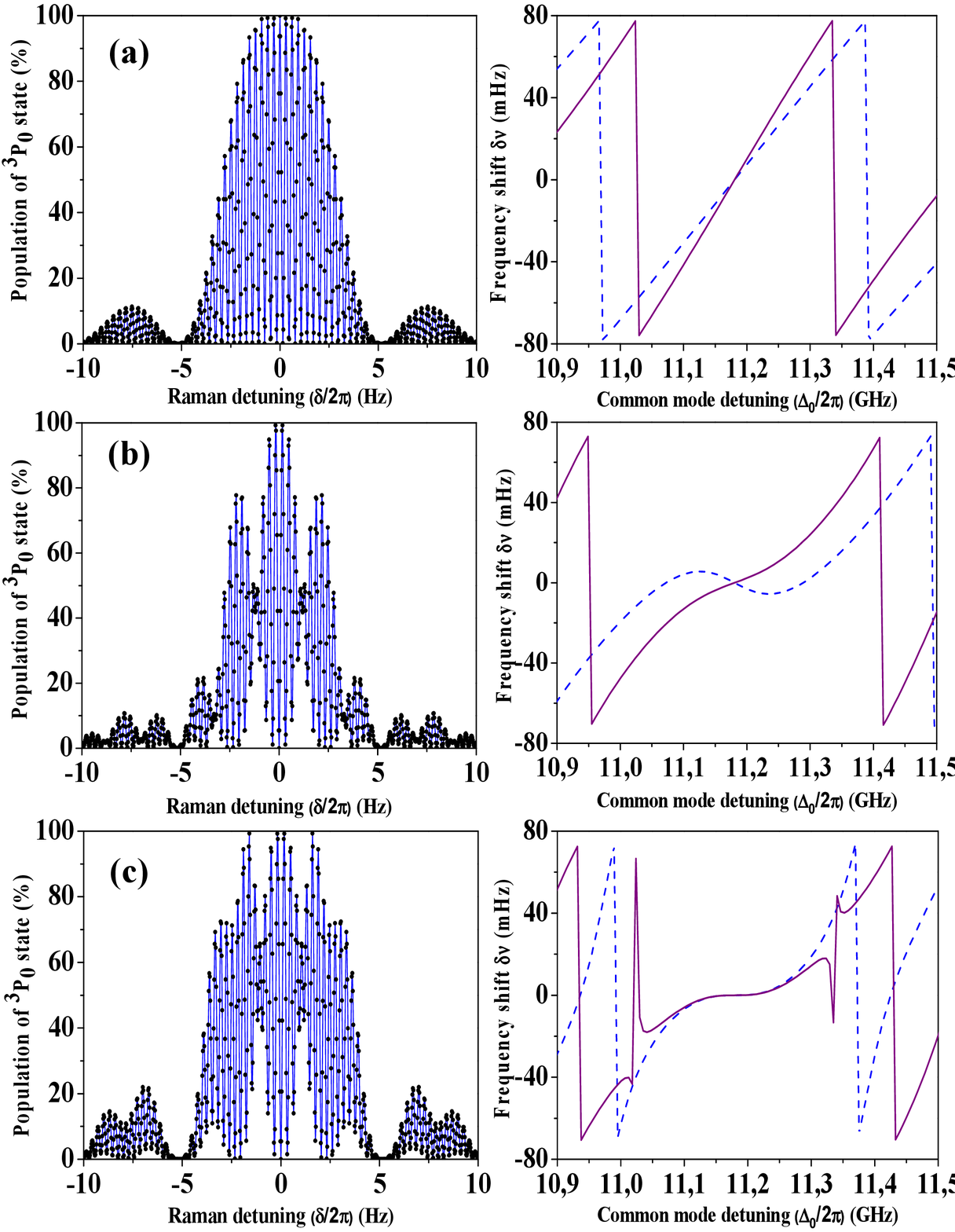}}
\caption{(color online) $^{88}$Sr transition probabilities measured on $^{3}P_{0}$ clock state (left panels) and $\delta\nu$ frequency-shifts versus the common mode detuning $\Delta_{0}$ (right panels) for different operating parameters all satisfying the primary condition $\delta \nu(\Delta_{0}^m)=0$. Rabi frequencies $\Omega_{1}=20\sqrt{\pi\Delta_{0}^{m}/3}$ and $\Omega_{2}=\Omega_{1}/100$, free evolution time $T=3$~s, pulse duration $\tau=0.1875=3/16$~s. In the right panel, the frequency-shifts are shown with a relative $\pm10\%$ pulse area variation (solid and dashed lines). All $\delta\nu(\Delta_0)$ curves computed from Eq.~(\ref{HER-phase-shift}). (a) Ramsey spectroscopy  based on the $\pi/2-T-\pi/2$ laser pulse sequence. (b) HRR scheme with the $\pi/2-T-3\pi/2$ pulse sequence. (c) HRR+Phase and Echo scheme with the $\pi/2-T-\pi-\pi/2$ pulse sequence including a phase reversal during the second pulse. In this case, a plateau is obtained around a magic common mode Raman detuning $\Delta_{0}^{m}/2\pi\sim11.2$~GHz, with all Eqs.~\eqref{conditions} conditions satisfied. }
\label{lightshifts}
\end{figure}
\begin{figure}[!!t]
\centering
\resizebox{8.0cm}{!}{\includegraphics[angle=0]{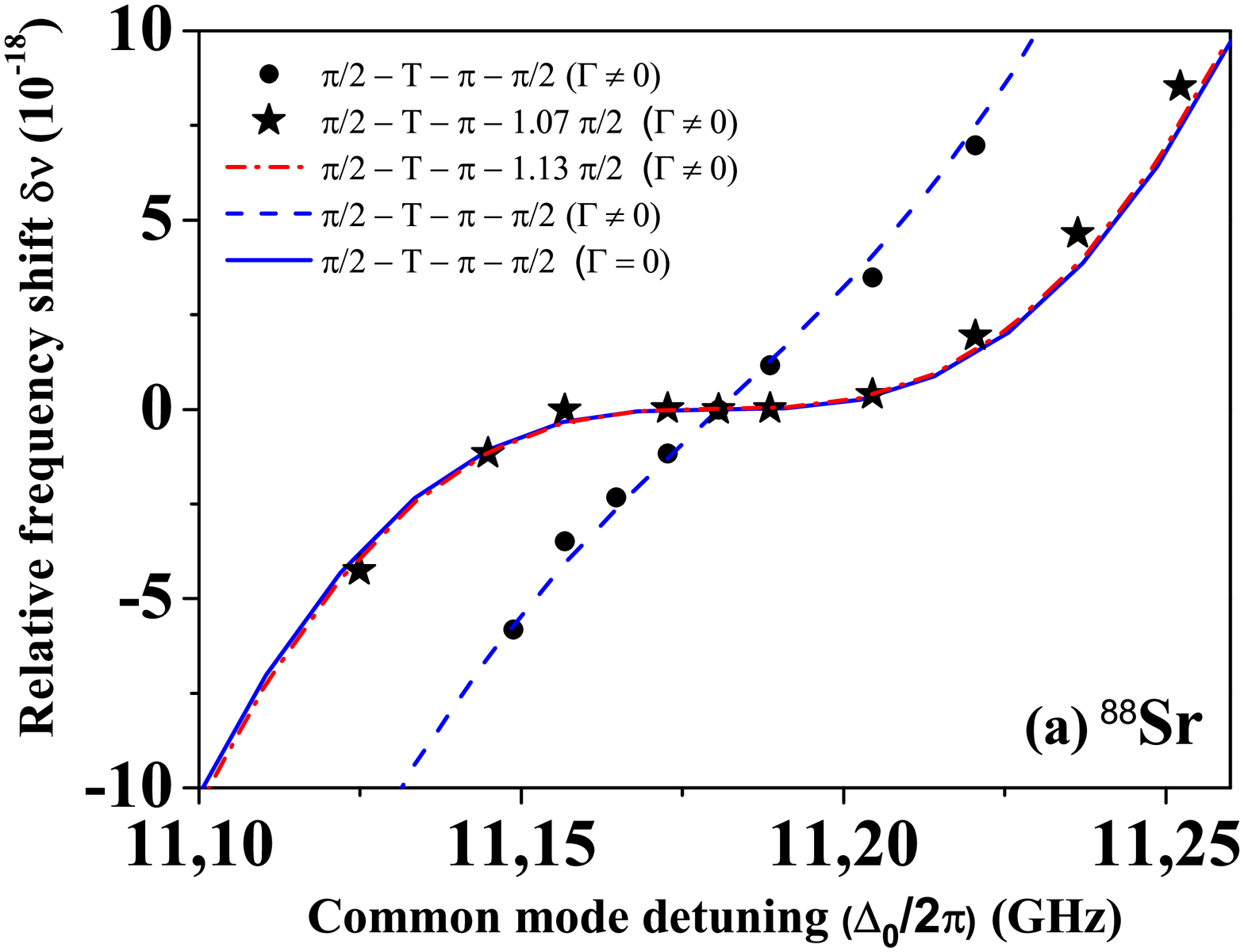}}
\resizebox{8.0cm}{!}{\includegraphics[angle=0]{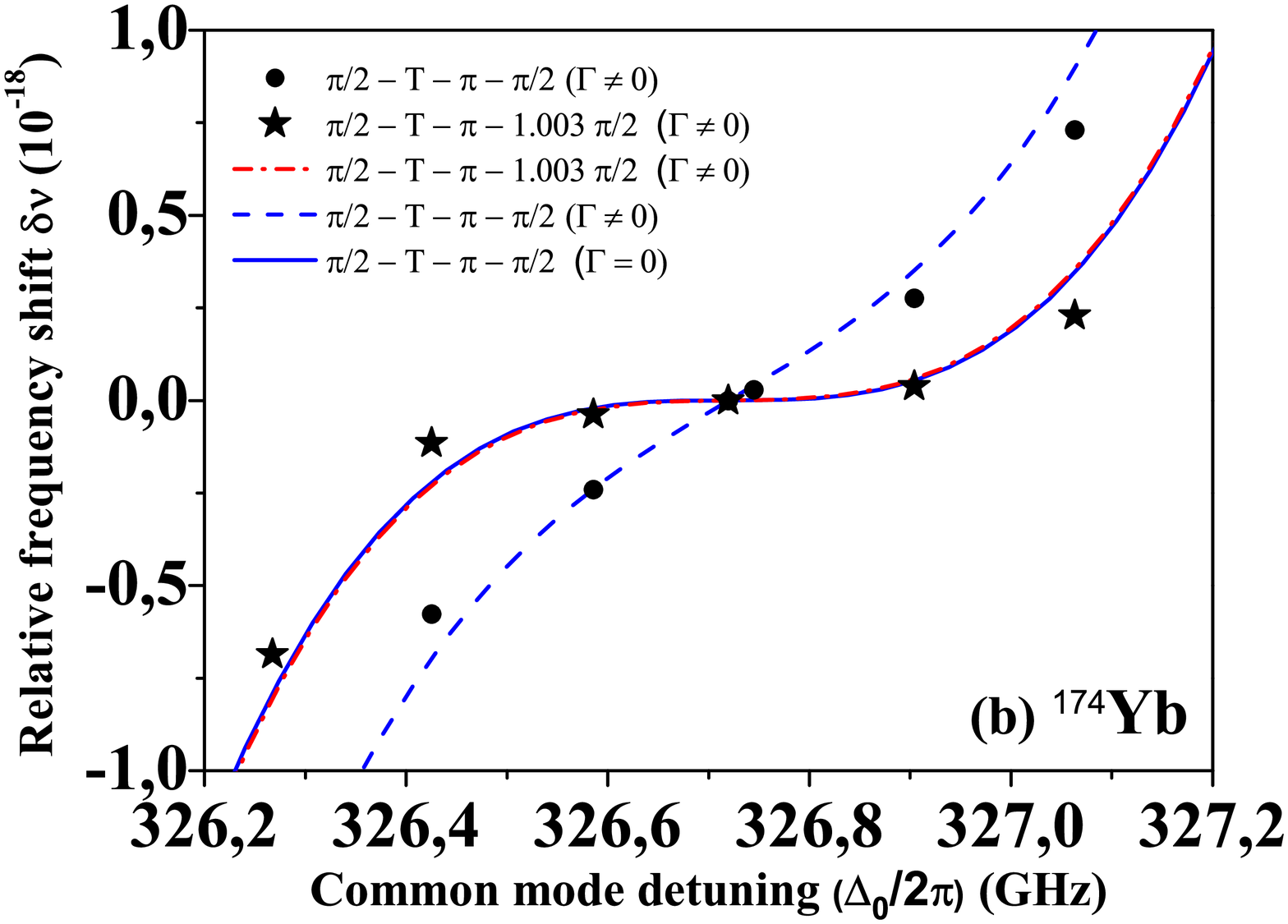}}
\caption{(Color online) Fractional frequency-shift (in $10^{-18}$ unit), based on a $\pi/2-T-3\pi/2$ laser pulse sequence, in the presence of spontaneous emission relaxation. All conditions from Eqs.~\eqref{conditions} are fulfilled at the mHz level, by adding between the preparation and detection pulses, an echo pulse of duration $\tau_{m}^{echo}=2\tau$, reversing the sign of one Rabi frequency.  (a) $^{88}$Sr clock frequency-shift around $\Delta_{0}^{m}=11.2$~GHz with $\Gamma/2\pi=32$~MHz and (b) $^{174}$Yb clock frequency-shift around $\Delta_{0}^{m}=326.7$~GHz  with $\Gamma/2\pi=28.9$~MHz. The density matrix analyzes (solid dots $\bullet$ and stars $\star$) are reproduced by the  complex wave-function model (dashed and solid  lines). Rabi frequencies as in Fig.~\ref{lightshifts}; free evolution time $T=5$~s. }
\label{clock-transition-Hyper-Raman-Sr-88}
\end{figure}
Different examples of shift cancelation with a fringe contrast nearly equal to one in the absence of spontaneous emission are presented in the left panels of Fig.~\ref{lightshifts}. Results for the synthesized $\delta \nu(\Delta_0)$ dependence to change in selected laser parameters are presented in the right panels of that figure  and in Fig.~\ref{clock-transition-Hyper-Raman-Sr-88} including spontaneous emission for different laser pulsed excitation schemes. The Ramsey sequence of Fig.~\ref{lightshifts}(a) with $\tau_m=\tau<<T$ and $\tau_{m}^{echo}\mapsto0$ produces the quasi linear $\delta \nu$ dependence on $\Delta_0=0$ of the right panel: only the $\delta \nu(\Delta_m^0)=0$ condition of Eqs.~\eqref{conditions} is satisfied. The HRR sequence of Fig.~\ref{lightshifts}(b) produces a $\delta\nu(\Delta_0)$ cubic dependence providing an excellent compensation and stability of the phase shift. The HRR+Phase scheme generalizing the two-level one of \cite{Yudin:2010}, where the laser pulse sequence includes a  phase reversal during the second pulse scheme produces a synthesized plateau against fluctuations in the laser frequency around the magic detuning, as in the right panel of Fig.~\ref{lightshifts}(c). Note that the associated fringes still have the maximum contrast. Magic detunings for other HRR interrogation parameters can be found in Tab.~\ref{Magic-fields-table}.\\
\indent Our synthesized light-shift approach allows the realization of shift cancelation with insensitivity to fluctuations of the laser intensity and/or pulse duration. For instance the HRR+Phase scheme suffers from an unstable shift compensation as a function of pulse area. That shift dependence on the pulse area is strongly eliminated within the synthesized phase shift approach using an additional echo pulse with phase reversal. The $\Omega_{eff}\to-\Omega_{eff}$ phase shift is introduced within the first part of the second pulse divided in two parts with areas $\pi$ and $\pi/2$, respectively. A very high  stability against laser intensity was verified.\\
\indent Figs.~\ref{lightshifts} and ~\ref{clock-transition-Hyper-Raman-Sr-88} examine clock interrogation with $\Delta_{0}^{m}$ frequency detunings in the GHz range,  therefore in a regime where the approximation of $\Gamma\mapsto0$ should be good enough. However a density matrix numerical simulation (fully confirmed by the wave-function model) pointed out that the spontaneous emission rate from the excited state cannot be totally neglected, as in~\cite{Tabatchikovaa:2013} for a two-level system.
The dots and dashed line HRR results of Fig.~\ref{clock-transition-Hyper-Raman-Sr-88} for $^{88}$Sr and $^{174}$Yb point out that at the magic detuning the slope $\partial\delta \nu/\partial \Delta_0$ is now different from zero, significantly compromising  the insensitivity of the cancelation to laser fluctuations. The relaxation introduces into the $\omega$ effective frequency a phase-shift modifying the coherent evolution of the atomic wave-function. The insensitivity is recovered by applying an HRR+Echo sequence with a finely tuned length of the third detection pulse, $\pi +1.07\pi/2$ for $^{88}$Sr at magic detuning $\Delta_{0}^{m}=11.2$~GHz, and $\pi +1.003\pi/2$ for $^{174}$Yb at $\Delta_{0}^{m}=326.7$~GHz, as shown by stars (density matrix) and continuous lines (wave-function model) in Figs.~\ref{clock-transition-Hyper-Raman-Sr-88}.

\section{Conclusion}

Our wave-function approach is very efficient for deriving the clock shift under different operating conditions and determining the laser parameters for the optical clock shift cancelation at the mHz level. We present its application to a Hyper-Raman-Ramsey spectroscopic method, based on two pulses with areas $\pi/2$ and $3\pi/2$, that cancels the light-shift and efficiently suppresses the sensitivity to laser field fluctuations without the additional frequency step on the laser frequency required in refs~\cite{Taichenachev:2009,Tabatchikovaa:2013,Huntemann:2012}.\\
\indent HRR spectroscopy has broader applications than light-shift cancelation control in clocks, focusing on the idea of quantum engineering of internal atomic and molecular states to prepare and probe complex quantum systems. Synthesized phase-shifts and operation at a magic Raman detuning could be applied to all high precision measurement using Ramsey spectroscopy, as in testing potential variation of fundamental constants with time \cite{Hudson:2006}, in high precision mass spectrometry with Penning traps \cite{Bollen:1992,Kretzschmar:2007} or in measuring gravitationally induced quantum phase shifts for neutrons~\cite{Abele:2010}. The light-insensitive two-photon approach could be applied also to molecular optical clocks sensitive to potential variation in the electron-to-proton mass ratio~\cite{Karr:2014}. In atomic interferometry, stimulated Raman transitions are intensively used to realise accurate inertial sensors, but still suffer from imprecise light-shift control~\cite{Gauguet:2014}. Our technique would be able to strongly suppress these systematic effects and could be more selective in the velocity class of atoms with less dispersion when including atomic recoil and Doppler effects. In time and frequency standards, there continues to be a  significant effort on pushing miniaturized microwave clock performance to higher levels, most of them employing a three-level coherent-population-trapping (CPT) interrogation. The HRR scheme may offer significant impact on all pulsed CPT/Raman clock designs~\cite{Vanier:2005,ShahKitching:2010} and  also in the case of the recently proposed E1-M1 portable optical clock ~\cite{Alden:2014}.

\section*{Acknowledgements}

\indent We gratefully acknowledge J. Lodewyck for providing Sr scalar polarizabilities to check the desired clock accuracy, K. Beloy for an estimate of magnetic dipole coupling in Yb, M. Glass-Maujean and C. Janssen for a careful reading of the manuscript.

\end{document}